\documentclass[10pt,english]{article}
\usepackage{graphicx}
\usepackage[T1]{fontenc}
\usepackage[latin9]{inputenc}
\usepackage[margin=1.0in]{geometry}
\usepackage{float}
\usepackage{amsmath}
\usepackage{amsbsy}
\usepackage{setspace}
\usepackage{amssymb}
\usepackage{esint}
\usepackage{cite}
\makeatletter

\floatstyle{ruled}
\newfloat{algorithm}{tbp}{loa}
\floatname{algorithm}{Algorithm}

\usepackage{babel}

\begin{document}

\title{Local Operators in Kinetic Wealth Distribution}

\author{M. Andrecut}

\date{March 3, 2016}

\maketitle
{

\centering Calgary, Alberta, T3G 5Y8, Canada

\centering mircea.andrecut@gmail.com

} 
\bigskip 
\begin{abstract}
The statistical mechanics approach to wealth distribution is based on the 
conservative kinetic multi-agent model for money exchange, where the local 
interaction rule between the agents is analogous to the elastic particle 
scattering process. Here, we discuss the role of a class of conservative local operators, 
and we show that, depending on the values of their parameters, they can be used 
to generate all the relevant distributions. We also show numerically that in order 
to generate the power-law tail an heterogeneous risk aversion model is required. 
By changing the parameters of these operators one can also fine tune the resulting 
distributions in order to provide support for the emergence of a more egalitarian 
wealth distribution.  
\end{abstract}
\bigskip 

\section{Introduction}

More than a century ago, Pareto has noticed that the wealth distribution in a 
stable economy seems to follow a power law distribution:
\begin{equation}
F(w)=\int_w^\infty f(w)dw\propto w^{-\nu}.
\end{equation}
where $F(w)$ is the cumulative probability of individuals with a wealth of at 
least $w$, and $f(w)$ is the probability distribution function \cite{key-1}. 
This power law, and the index $\nu$ are known today as Pareto law, and 
respectively Pareto index. 

Further studies have shown that Pareto's law only explains the distribution of 
the higher income class, situated on the tail of the wealth distribution. 
Away from the tail, the distribution is better described by a Gamma or Log-normal 
distribution known as Gibrat's law \cite{key-2}. More recent studies also 
suggest that the low income class can be described by Boltzmann-Gibbs and Gamma-like 
distributions $f_{G}(w)$ \cite{key-3,key-4,key-5,key-6,key-7,key-8,key-9,key-10}:
\begin{align}
 f(w) &\propto 
  \begin{cases}
   f_{G}(w)  & \text{for } w < w_c \\
   w^{-\nu-1} & \text{for } w\geq w_c
  \end{cases}.
\end{align}
These recent studies show that the Pareto law with an index $\nu \in [1,2]$, 
with slight variations, exhibits a remarkable stability in the real data, 
showing that a small percentage of people are holding the majority of wealth. 

Several multi-agent models of closed economies have been proposed to explain 
these empirical results (for a review see \cite{key-11} and the references 
therein). These models borrow methods from the statistical mechanics of 
particle systems, and their basic assumption is that wealth is exchanged among 
agents at a microscopic level, using a simple pairwise trade mechanism, which is 
physically equivalent to a particle scattering process \cite{key-12}. 
Despite their simplicity, these models exhibit the characteristics of real 
data and provide a framework for better understanding the complexity behind 
the wealth distribution mechanism. 

Most of these models are based on conservative microscopic interactions, 
and have focused on incorporating a saving mechanism, which 
allows only a fraction of the agent's wealth to participate in each trading event \cite{key-13,key-14}.
More recently, effort was made to include macroscopic 
mechanisms for global taxation and wealth redistribution in an effort to 
counter the concentration of all available wealth in few agents, and to 
favour the poorer agents \cite{key-15}. These mechanisms can also lead to the 
creation of a middle class, since the models implementing them exhibit a 
transition from the "unfair" Boltzmann-Gibbs distribution to a Gamma-like unimodal 
distribution, where most of the agent's wealth is shifted away from zero to a 
positive value, corresponding to the mode of the distribution. 

Here, we discuss the role of a class of conservative local operators, 
and we show that, depending on the values of their parameters, they can be used 
to generate all the relevant wealth distributions:  
 Dirac delta, Boltzmann-Gibbs, Gamma and Pareto.
We also show numerically that in order to generate the power-law tail 
an heterogeneous risk aversion model is required. 
By changing the parameters of these operators one can also fine tune the 
resulting distributions in order to provide support for the emergence of 
a more egalitarian wealth distribution. 

\section{Kinetic models for wealth distribution}
While wealth is a complex concept that includes money, 
property and other material goods that have a certain economic utility, here we 
consider that wealth is measured only in terms of money. We also assume that 
money is the exchange medium used in economic transactions between agents, and 
that the total amount of money is a conserved quantity in a closed economy. 

We consider a closed economy with $N$ agents, characterized by their wealth 
state $w_n \geq 0$, $n=1,2,...,N$. The economic transactions are described 
as binary interactions, where at a given time step, two random agents $i$ and 
$j$ are exchanging an amount of wealth (money) $\Delta w$ between them:
\begin{align}
w'_i &= w_i - \Delta w, \nonumber \\
w'_j &= w_j + \Delta w,
\end{align}
such that total amount of money of the two agents, before and after transaction, 
is conserved:
\begin{equation}
w'_i + w'_j = w_i + w_j.
\end{equation}
This local conservation law for money is analogous to the energy conservation 
in the elastic collisions between the molecules of an ideal gas. 

In the most closely related model to traditional statistical mechanics the 
transaction wealth exchange quantity is defined as:
\begin{equation}
\Delta w = (1-\eta)w_i - \eta w_j,
\end{equation}
such that the local transaction equations are:
\begin{align}
w'_i &= \eta (w_i + w_j) \nonumber, \\
w'_j &= (1 - \eta) (w_i + w_j),
\end{align}
where $\eta \sim U[0,1]$ is a uniform distributed random variable, 
governing the random redistribution of the wealth of the agents \cite{key-11}. 
It has been shown that this basic model leads to an equilibrium Boltzmann-Gibbs 
distribution:
\begin{equation}
f(w) = \frac{1}{\langle w \rangle} 
		\exp \left( -\frac{w}{\langle w \rangle} \right),
\end{equation}
with the "temperature" equal to the average amount of money per agent 
$\langle w \rangle$ \cite{key-12}. This result is extremely robust and 
independent of various factors, such as arbitrary initial conditions and 
random or consecutive interaction of agents. The Boltzmann-Gibbs distribution is 
characterized by very few rich agents and a majority of poor agents, 
without a well defined middle class. 

In order to overcome the "unfairness" of the Boltzmann-Gibbs wealth distribution, 
one can include a savings mechanism that leads to a more "fair" 
Gamma-like equilibrium distribution \cite{key-13}. 
In this case the agents save some fraction of their money $\lambda w_i$, 
and use the rest of their money balance, $(1-\lambda)w_i$, for random 
exchanges, such that the locally conservative transaction equations become:
\begin{align}
w'_i &= \lambda w_i + \eta (1-\lambda)(w_i + w_j), \nonumber \\
w'_j &= \lambda w_j + (1 - \eta) (1-\lambda) (w_i + w_j).
\end{align}
The coefficient $\lambda \in [0,1]$ is a global constant called the saving 
propensity. It was also shown that the equilibrium distribution of this 
model is close to a Gamma distribution \cite{key-13}.

A further enhancement of this model considers different saving propensities 
for the agents.\cite{key-14} The propensities also maintain their 
values fixed during the relaxation dynamics of the system. 
It was found that this model produces a Gamma-like distribution ending with 
a Pareto tail with an index $\nu\simeq1$, which originates from the agents 
with a saving propensity close to one \cite{key-14}.

A different approach considers a macroscopic mechanism for global taxation 
and wealth redistribution which favours the poorer agents \cite{key-15}. 
In such a model, any transaction can be described in two steps. In the first 
step, two random agents $i$ and $j$ exchange their money such that a fraction 
$f \in [0,1)$ of their exchanged wealth is lost by taxes:
\begin{align}
w_i(t+1/2) &= (1-f)\eta [w_i(t) + w_j(t)], \nonumber \\
w_j(t+1/2) &= (1-f)(1 - \eta) [w_i(t) + w_j(t)].
\end{align}
In the second step, the collected taxes $\tau = f[w_i(t) + w_j(t)]$ are 
equally redistributed to all the agents in the population:
\begin{equation}
w_n(t+1) = w_n(t+1/2) + \tau /N, \: n=1,...,N.
\end{equation}
The transactions can be described by inelastic collisions between gas 
particles, where the energy lost is the analogue of the taxation value. 
The model exhibits a transition from the Boltzmann-Gibbs distribution to a Gamma-like 
distribution as $f$ increases, and after a critical point it returns to an 
exponential for higher $f$ values.

\section{Local operators in kinetic wealth distribution}
Here we discuss the role of a class of conservative local operators 
in the wealth exchange dynamics. Also, in order to characterize the 
wealth distributions we use the mode of the 
distribution, when it is present, and the Gini coefficient, 
which measures the inequality among the values in the distribution \cite{key-16}. 
A Gini coefficient $G=1$ expresses the maximal wealth inequality, while a Gini 
coefficient $G=0$ corresponds to perfect equality, where all the wealth 
values are the same. 
For a population $w_n \geq 0$, $n=1,2,...,N$, that is indexed in non-decreasing 
order ($w_n \leq w_{n+1}$), the Gini coefficient can be easily computed as 
following \cite{key-17,key-18}:
\begin{equation}
G = \frac{1}{N}\left[ N+1 - 2 \frac{\sum_{n=1}^N (N+1-n)w_n}
	{\sum_{n=1}^N w_n} \right]
\end{equation}
Also, in all the computations we impose a conserved total wealth: 
\begin{equation}
W = \sum_{n=1}^N w_n = N
\end{equation}

Let us first consider the basic model described by the equation (6), written 
in the following matrix form:
\begin{equation}
\begin{bmatrix}
    w_i(t+1)\\
    w_j(t+1)
\end{bmatrix}
=T^{+}(\varepsilon)
\begin{bmatrix}
    w_i(t)\\
    w_j(t)
\end{bmatrix}
\end{equation}
where 
\begin{equation}
T^{+}(\varepsilon) = 
\begin{bmatrix}
    \varepsilon & \varepsilon  \\
    1 - \varepsilon & 1 - \varepsilon 
\end{bmatrix}
\end{equation}
and $\varepsilon \sim U[0,1]$ is an uniformly distributed random variable.
One can see that this is a stochastic operator corresponding to a column 
stochastic matrix, which is also singular, and therefore non-invertible.  
It is interesting to note that 
$T^{+}(\varepsilon)$ has no memory of its previous applications:
\begin{equation}
T^{+}(\varepsilon)T^{+}(\varepsilon') = T^{+}(\varepsilon), 
\:\: \forall \varepsilon' \neq \varepsilon
\end{equation}
which means that the result of its successive applications only depends on the 
last application \cite{key-19}. Also, it is well known that starting from any 
distribution of money, and by applying this local interaction rule, the steady 
state of the wealth distribution becomes a Boltzmann-Gibbs distribution \cite{key-12}. 

Let us now replace the $T^{+}(\varepsilon)$ operator in this basic model with 
the following operator:
\begin{equation}
T^{-}(\varepsilon) = 
\begin{bmatrix}
    \varepsilon & 1 - \varepsilon  \\
    1 - \varepsilon & \varepsilon 
\end{bmatrix}
\end{equation}
where $\varepsilon \sim U[0,1]$. This operator corresponds to a non-singular 
double stochastic matrix, which is not memoryless, 
since its successive applications accumulate as follows:
\begin{align}
T^{-}(\varepsilon)T^{-}(\varepsilon') = T^{-}(\varepsilon'') \nonumber \\
T^{-}(\varepsilon)T^{+}(\varepsilon') = T^{+}(\varepsilon'') 
\end{align}
where 
$\varepsilon''= \varepsilon \varepsilon' + (1-\varepsilon)(1-\varepsilon')$.

\begin{figure}[!ht]
\centering \includegraphics[width=16.5cm]{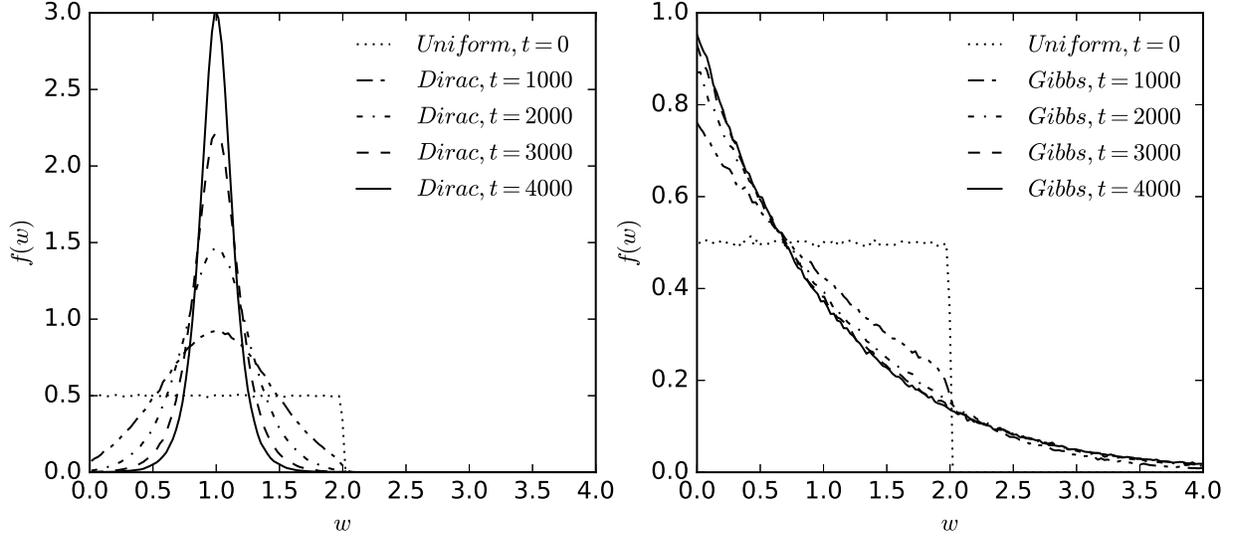}\caption{Time evolution of 
a uniform wealth distribution into a Dirac delta function (left) and a Boltzmann-Gibbs 
distribution (right), using the $T^{-}(\varepsilon)$ and respectively $T^{+}(\varepsilon)$ 
operators.}
\end{figure}

Our numerical simulations show that by applying this local interaction rule, 
the steady state of any wealth distribution becomes a Dirac delta function, 
independent of the initial distribution. 
In Figure 1 we show such a simulation, where a population of $N=1000$ agents 
with an initial wealth $W=N$, uniformly distributed among them in $U[0,2]$ with 
an average $\langle w \rangle = 1$, is evolved to 
either a Dirac delta function (Fig.1, left), or a Boltzmann-Gibbs distribution 
(Fig.1, right). During the numerical simulation, a pair of agents is randomly 
drawn at every time step, and they interact locally using the 
$T^{-}(\varepsilon)$ or $T^{+}(\varepsilon)$ operators, respectively. 
After about $10^6$ steps, the initial uniform distribution is gradually 
transformed into a Dirac delta function $f(w)\sim\delta(w-1)$, where all the agents have 
an identical wealth $w_n=1, n=1,...,N$, or an exponential Boltzmann-Gibbs distribution 
$f(w)\sim exp(-w)$, with the "temperature" $\langle w\rangle=1$. The results shown in 
Figure 1 are for $100$, $2000$, $3000$ and $4000$ time steps, averaged over 
$10^4$ initial configurations.

These two operators $T^{+}(\varepsilon)$ and $T^{-}(\varepsilon)$ can be used 
to transform any given uniform distribution into a Boltzmann-Gibbs and respectively a 
Dirac distribution, which also means that they can be used to switch back and 
forth between these two extreme distributions: 
\begin{equation}
Dirac\:delta \; \xleftarrow {\:T^-\:}\; Uniform\; \xrightarrow {\:T^+\:} \; Boltzmann-Gibbs
\end{equation}
Thus, the $T^{-}(\varepsilon)$ and $T^{+}(\varepsilon)$ operators exhibit an 
antagonistic behaviour, since $T^{-}(\varepsilon)$ can be used to create a perfectly   
egalitarian distribution (Dirac), while $T^{+}(\varepsilon)$ is responsible 
for creating an unfair distribution (Boltzmann-Gibbs).  

One can easily create other operators by interpolating these two operators 
using a new variable $\xi \in [0,1]$: 
\begin{equation}
T^{\pm}(\varepsilon, \varepsilon',\xi) = 
\xi T^{-}(\varepsilon) + (1 - \xi) T^{+}(\varepsilon')
\end{equation}
Obviously, by mixing two conservative operators, the resulting operator will 
also satisfy the local money conservation rule. This operator allows us to 
fine tune the distribution between the two extremes corresponding to Dirac delta 
and Boltzmann-Gibbs distributions. 

\begin{figure}[!ht]
\centering \includegraphics[width=16.5cm]{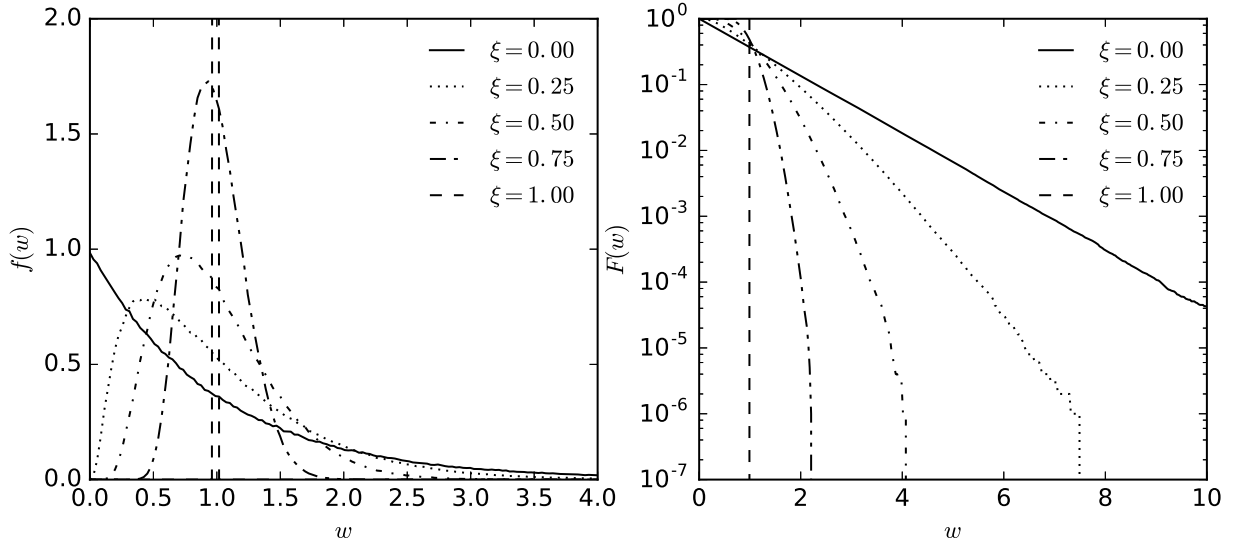}
\caption{Probability distribution function $f(w)$, and cumulative distribution 
function $F(w)$, for the interpolating operator 
$T^{\pm}(\varepsilon, \varepsilon',\xi)$, and different values of $\xi$.}
\end{figure}

In Figure 2 we show the results of the simulation for 
$\xi=0.00$, $0.25$, $0.50$, $0.75$, $1.00$. The wealth distribution that results by mixing 
these these two antagonistic operators is similar to a Gamma distribution. Here 
we have considered a population of $N=1000$ agents with an initial wealth $W=N$, 
uniformly distributed among them. The number of simulation steps was set to 
$10^6$, and we averaged over $10^4$ initial configurations. The results for 
the Gini coefficient, $G(\xi)$, and the first mode, $\mu$, of the distribution 
are also shown Table 1.

\bigskip 
\begin{center}
\begin{tabular}{|l|r|r|r|r|r|}
\hline 
$\:$ & $\xi=0.00$ & $\xi=0.25$ & $\xi=0.50$ & $\xi=0.75$ & $\xi=1.00$ \\ 
\hline 
$G$ & 0.500 & 0.365 & 0.246 & 0.129 & 0.000 \\ 
\hline 
$\mu$ & 0.000 & 0.454 & 0.776 & 0.922 & 1.000 \\ 
\hline 
\end{tabular} 
\end{center}
\bigskip

\noindent
The Gini coefficient takes the maximum value $G=0.5$ for $\xi=0$, 
corresponding to the Boltzmann-Gibbs distribution, and gradually decreases to the minimum 
value $G=0$, corresponding to the Dirac distribution, as the parameter $\xi$ 
approaches $1$. The mode $\mu$ of the Gamma-like distribution increases with 
$\xi$, from $0$ (Boltzmann-Gibbs) to $1$ (Dirac delta).

The above operators can be generalized by considering that the money exchange 
between the agents is governed by two independent uniform random variables, 
$\varepsilon, \rho \sim U[0,1]$,  
such that the local conservation rule is still respected. Thus we have:
\begin{equation}
T^{+}(\varepsilon, \rho) = 
\begin{bmatrix}
    \varepsilon & \rho  \\
    1 - \varepsilon & 1 - \rho
\end{bmatrix}
\end{equation}
and respectively:
\begin{equation}
T^{-}(\varepsilon, \rho) = 
\begin{bmatrix}
    \varepsilon & 1 - \rho  \\
    1 - \varepsilon & \rho
\end{bmatrix}
\end{equation}
One can see that both operators are column stochastic matrices, and they are both 
invertible for $\varepsilon \neq \rho$, which suggests a different behaviour.

\begin{figure}[!ht]
\centering \includegraphics[width=16.5cm]{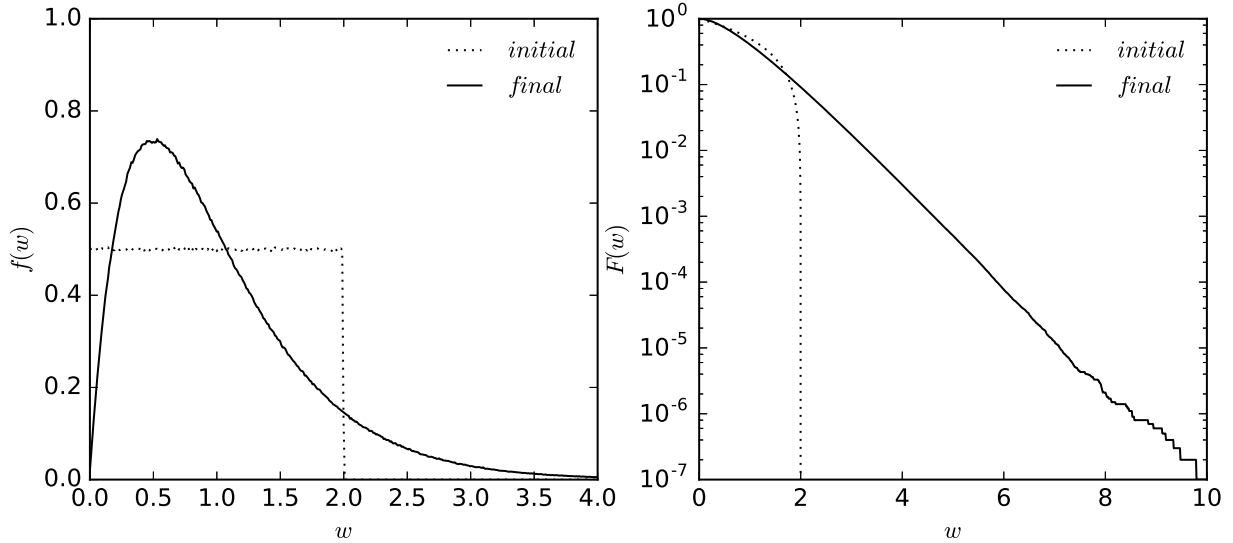}\caption{Probability 
distribution function $f(w)$, and cumulative distribution function $F(w)$, 
for the statistically equivalent operators 
$T^{+}(\varepsilon, \rho) \equiv T^{-}(\varepsilon, \rho)$.}
\end{figure}

The result of these operators is obviously different when they act locally on a pair of 
agents. However, their "global" statistical behaviour is the same if $\varepsilon$ and 
$\rho$ are independent uniform distributed random variables, sampled at every iteration step:
\begin{equation}
T^{+}(\varepsilon, \rho) \equiv T^{-}(\varepsilon, \rho)
\end{equation}
which means that they are "statistically equivalent".

Our numerical simulations show that these operators transform any initial 
distribution into a Gamma distribution: 
\begin{equation}
f(w) = \frac{w^{k-1}e^{-w/\theta}}{\Gamma (k) \theta^k}
\end{equation}
with $k=2$, $\theta=2$, and a Gini coefficient:
 \begin{equation}
G = \frac{\Gamma(5/2)}{2\Gamma(2)\sqrt{\pi}}=0.375
\end{equation}
In Figure 3 we give such an example, where the 
initial distribution corresponds to a population of $N=1000$ agents with an 
initial wealth $W=N$, uniformly distributed among them. The number of 
simulation steps was set to $10^6$, and we averaged over $10^4$ initial 
configurations. 

Let us now consider a heterogeneous model, where the agents have a different "personality", 
which is described by the following operator:

\begin{equation}
T^{+}(\varepsilon_i, \rho_j) = 
\begin{bmatrix}
    \varepsilon_i & \rho_j  \\
    1 - \varepsilon_i & 1 - \rho_j
\end{bmatrix}
\end{equation}
Here, the parameters $\varepsilon_n, \rho_n, n=1,...,N$ can be interpreted 
as the risk aversion of the agents, and they are initially drawn 
from a uniform random distribution $U[0,1]$, and contrary to the previous models 
they are kept frozen during the computation. 

\begin{figure}[!ht]
\centering \includegraphics[width=16.5cm]{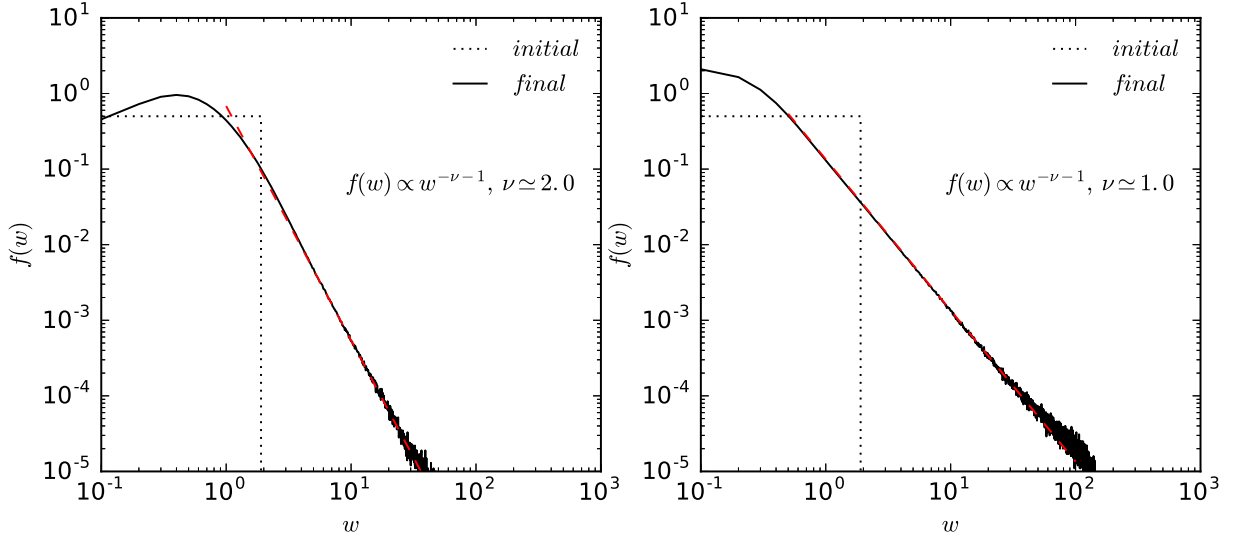}\caption{Pareto distribution with 
index $\nu \simeq 2$ for the operator $T^{+}(\varepsilon_i, \rho_j)$, with fixed 
values $\varepsilon_i, \rho_j \sim U[0,1]$, $i,j=1,...,N$ (left); and the 
Pareto law with index $\nu \simeq 1$ for the heterogeneous savings model (right). 
The dashed red line is the fit with the Pareto distribution.}
\end{figure}

The obtained result is quite different, and it shows a unimodal distribution 
with a Pareto tail with the index $\nu\simeq 2$, Figure 4 (left). The Gini 
coefficient of the distribution in this example is $G = 0.447$. 
The dashed red line in Figure 4 is the fit with the Pareto distribution.

For comparison we have also considered the heterogeneous savings model, 
which generates a Pareto distribution with an 
index $\nu \simeq 1$, and a much larger Gini coefficient $G = 0.759$, 
Figure 4 (right) \cite{key-14}. Thus the distribution generated by the operator 
$T^{+}(\varepsilon_i, \rho_j)$ is more "fair" than the distribution 
generated by the heterogeneous savings model, indicating more wealth 
accumulated by the middle class.

The explanation for this difference is that the parameters in the models are 
different. The heterogeneous savings model can be written as following \cite{key-14}:
\begin{align}
w'_i &= \lambda_i w_i + \eta [(1-\lambda_i)w_i + (1-\lambda_j)w_j] 
\nonumber \\
w'_j &= \lambda_j w_j + (1 - \eta) ((1-\lambda_i)w_i + (1-\lambda_j)w_j)
\end{align}
where $\lambda_i \sim U[0,1]$ are the fixed saving propensities of the agents, 
and $\eta \in U[0,1]$ is a uniform random variable.
We notice that this model can be also rewritten in a matrix form as:
\begin{equation}
\begin{bmatrix}
    w'_i\\
    w'_j
\end{bmatrix}
=T^{+}(\varepsilon'_i(\eta,\lambda_i),\rho'_j(\eta,\lambda_j))
\begin{bmatrix}
    w_i\\
    w_j
\end{bmatrix}
\end{equation}
with:
\begin{align}
\varepsilon'_i (\eta,\lambda_i) &= 1 - (1 - \eta) (1-\lambda_i) \nonumber \\
\rho'_j (\eta,\lambda_j) &= (1 - \eta) (1-\lambda_j)
\end{align}
Thus, the models have similar equations, and we could have expected to obtain 
similar results. However, the results are quite different due to the difference 
in the definition of their parameters. 

\begin{figure}[!ht]
\centering \includegraphics[width=16.5cm]{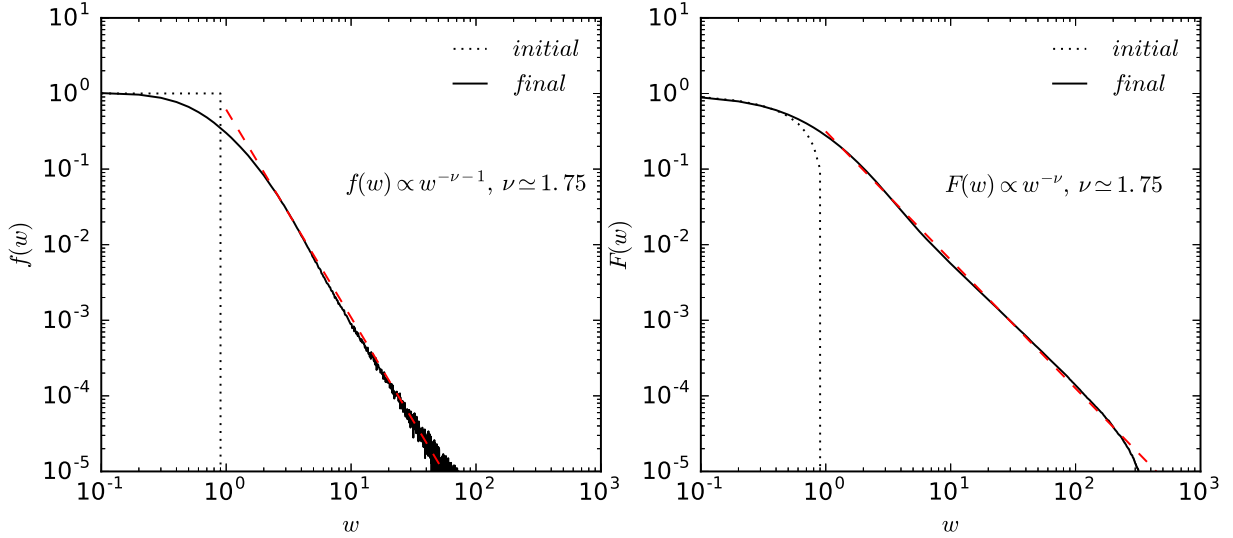}\caption{Pareto distribution with 
index $\nu \simeq 1.75$ for the operator $T^{+}(\varepsilon_i, \rho_j)$, with 
fixed values $\varepsilon_i, \rho_j \in (0,1)$, $i,j=1,...,N$ drawn from the 
square of a uniform random distributed variable. The dashed red line is the fit with the Pareto distribution.}
\end{figure}

The parameters $\varepsilon_i$, $\rho_j$ 
in the operator $T^{+}(\varepsilon_i, \rho_j)$ are initially drawn from a uniform 
random distribution $U[0,1]$, and then kept frozen during the simulation, while 
the parameters $\varepsilon'_i$ and $\rho'_j$ in the heterogeneous savings 
model correspond to the product of a uniform distributed random variable 
$x\equiv 1-\eta$, changing randomly at every step, and 
$y\equiv 1-\lambda_j$ which is also kept frozen. 

The index of the Pareto law can be further modified by changing the 
distribution of the parameters $\varepsilon_i$ and $\rho_j$ in $T^{+}(\varepsilon_i, \rho_j)$. 
For example, if $\varepsilon_i$ and $\rho_j$ are drawn from $1-\xi^2$, where $\xi$ is a uniform random variable $\xi\sim U[0,1]$, 
then the index becomes $\nu\simeq 1.75$, as show in Figure 5. The Gini coefficient for the distribution in this 
example also increases to $G = 0.596$. 

The changes in the Pareto index can 
be explained by the fact that the density function of $1-\xi^2$ is not  
uniform any more. The density function of $1-\xi^2$ can be easily calculated by 
first calculating the cumulative distribution as follows:
\begin{align}
F_{1-\xi^2}(\xi) &= \text{Pr}(1-\xi^2\leq x) \nonumber \\
&= \text{Pr}(x \in [\sqrt{1-\xi},1]) \nonumber \\
&=1 - \sqrt{1- \xi}
\end{align}
Thus, the density function of $1-\xi^2$ is:
\begin{equation}
f_{1-\xi^2}(\xi) = \frac{dF_{1-\xi^2}(\xi)}{d\xi} = \frac{1}{2 \sqrt{1-\xi}}
\end{equation} 
A smaller Pareto index $\nu\simeq 1.6$ is obtained if the parameters $\varepsilon_i$ and $\rho_j$ are drawn from $1-\xi^4$, where $\xi\sim U[0,1]$.   
This means that the Pareto index decreases by further increasing the probability of higher risk aversion parameters.

Therefore, the index of the Pareto distribution in these models 
strongly depends on the distribution of the parameters $\varepsilon_i$ and $\rho_j$ 
in the operator $T^{+}(\varepsilon_i, \rho_j)$. 

\section{Conclusion}
We have discussed the role of a class of conservative local operators, that 
can be used to reproduce the main features of empirical distributions observed in the 
previous kinetic models of wealth distribution: Boltzmann-Gibbs, Dirac delta, Gamma and Pareto.
Numerical simulations have shown that in order to generate the power-law tail 
an heterogeneous risk aversion model is required. In this case, the Pareto index 
strongly depends on the distribution of the risk aversion parameters. 
Also, we have shown numerically that by changing the parameters of these operators 
one can also fine tune the resulting distributions in order to provide support for 
the emergence of a more egalitarian wealth distribution.  

As a closing remark, we would like to note that these results can be further extended to more complex 
agent models in econophysics and sociophysics, where a binary exchange mechanism between 
the agents is expected (market models, opinion formation etc.). Also, this class of local 
conservative operators can be applied to more relaxed cases where the local 
exchanged quantities are only conserved in the mean:
\begin{equation}
w'_i+w'_j=\langle w_i+w_j \rangle,
\end{equation}
removing the strong pointwise conservation constraint, and allowing for a more complex stochastic 
dynamics where other mechanisms like debt and interest can be taken into account.


\begin{thebibliography}{99}

\bibitem{key-1} V. Pareto, 
				{\it Cours d'economie Politique}  
				(F. Rouge, Lausanne, 1897).
\bibitem{key-2} Gibrat, R., 
				{\it Les In\'{e}galit\'{e}s Economiques Sirely} (Paris, 1931).
\bibitem{key-3} S. Moss de Oliveira, P.M.C. de Oliveira and D. Stauffer, 
				{\it Evolution, Money, War and Computers} (B.G. Tuebner, Stuttgart, Leipzig, 1999).
\bibitem{key-4} M. Levy, S. Solomon, 
				{\it Physica A} 242, 90 (1997).
\bibitem{key-5} H. Aoyama, W. Souma, Y. Fujiwara, 
				{\it Physica A} 324, 352 (2003)
\bibitem{key-6} F. Clementi, M. Gallegati, 
				{\it Physica A} 350, 427 (2005)
\bibitem{key-7} A. Banerjee, V.M. Yakovenko, T.Di Matteo, 
				{\it Physica A} 370, 54 (2006).
\bibitem{key-8} S. Sinha, 
				{\it Physica A} 359, 555, (2006).
\bibitem{key-9} A.A. Dragulescu, V.M. Yakovenko,
				{\it Eur. Phys. J. B} 20, 585 (2001)
\bibitem{key-10} A.A. Dragulescu, V.M. Yakovenko, 
				{\it Physica A} 299, 213 (2001)
\bibitem{key-11} V.M. Yakovenko, J. Barkley Rosser Jr., 
				{\it Rev. Mod. Phys.} 81, 1703 (2009)
\bibitem{key-12} A.A. Dragulescu, V.M. Yakovenko, 
				{\it Eur. Phys. J. B} 17, 723 (2000)
\bibitem{key-13} A. Chakraborti, B.K. Chakrabarti, 
				{\it Eur. Phys. J. B} 17, 167 (2000).
\bibitem{key-14} A. Chatterjee, B.K. Chakrabarti, S.S. Manna, 
				{\it Physica A} 335, 155 (2004).
\bibitem{key-15} S. Guala, 
				{\it Int. Desc. Comp. Sys.} 7, 1 (2009)
\bibitem{key-16} C. Gini, 
				{\it Rivista di Politica Economica} 87, 769 (1997).
\bibitem{key-17} P.M. Dixon, J. Weiner, T. Mitchell-Olds, R. Woodley, 
				{\it Ecology} 68, 1548 (1987). 
\bibitem{key-18} C. Damgaard, J. Weiner, 
				{\it Ecology} 81, 1139 (2000).
\bibitem{key-19} A.K. Gupta, 
				{\it Physica} A 359, 634 (2006).

\end{thebibliography}
\end{document}